\documentclass[aps,prl,twocolumn,showpacs]{revtex4}
\usepackage{graphicx}
\usepackage{times}
\usepackage{latexsym}
\def\beq{\begin{equation}}
\def\eeq{\end{equation}}
\def\bey{\begin{eqnarray}}
\def\eey{\end{eqnarray}}

\def\lsim{\mathrel{\raise.3ex\hbox{$<$\kern-.75em\lower1ex\hbox{$\sim$}}}}
\def\gsim{\mathrel{\raise.3ex\hbox{$>$\kern-.75em\lower1ex\hbox{$\sim$}}}}

\newcommand{\be}{\begin{equation}}
\newcommand{\ee}{\end{equation}}

\newcommand{\la}{\langle}
\newcommand{\ra}{\rangle}

\newcommand{\MeV}{{\rm ~MeV }}

\begin{document}

\title{A Natural Supersymmetric Model with MeV Dark Matter}  
\author{Dan Hooper$^1$ and Kathryn M. Zurek$^2$}
\address{$^1$Theoretical Astrophysics, Fermi National Accelerator Laboratory, Batavia, IL  60510 \\ $^2$Physics Department, University of Wisconsin, Madison, WI  53706}

\date{\today}

\begin{abstract}

It has previously been proposed that annihilating dark matter particles with MeV-scale masses could be responsible for the flux of 511 keV photons observed from the region of the Galactic Bulge. The conventional wisdom, however, is that it is very challenging to construct a viable particle physics model containing MeV dark matter. In this letter, we challenge this conclusion by describing a simple and natural supersymmetric model in which the lightest supersymmetric particle naturally has a MeV-scale mass and the other phenomenological properties required to generate the 511 keV emission. In particular, the small ($\sim$\,$10^{-5}$) effective couplings between dark matter and the Standard Model fermions required in this scenario naturally lead to radiative corrections that generate MeV-scale masses for both the dark matter candidate and the mediator particle.

\end{abstract}
\pacs{95.35.+d; 95.30.Cq; 12.60.Jv; FERMILAB-PUB-07-587-A; MADPH-07-1499}
\maketitle

In 2003, it was announced that the SPI spectrometer on board the INTEGRAL satellite had confirmed the presence of a very bright flux of 511 keV photons from the region of the Galactic Bulge. This emission corresponds to approximately $3 \times 10^{42}$ positrons being injected per second in the inner kiloparsecs of the Milky Way~\cite{integral}. The signal is approximately spherically symmetric (with a full-width-half-maximum of 6$^{\circ}$), with little of the emission tracing the Galactic Disk. The origin of these positrons remains unknown.

It is somewhat difficult to explain the observed 511 keV emission with astrophysical mechanisms. Firstly, it is not clear that known astrophysical sources are able to inject a large enough number of positrons to generate the observed signal. In particular, type Ia supernovae cannot produce enough positrons to generate the observed flux~\cite{Kalemci:2006bz}. In the case of either hypernovae~\cite{Casse:2003fh}, gamma ray bursts~\cite{Casse:2003fh,Parizot:2004ph} or microquasars~\cite{microquasars}, this is less clear. In order for hypernovae or gamma rays bursts to generate the observed intensity of the 511 keV emission, they would have to occur rather frequently within the inner Milky Way (approximately 0.02 per century or 0.0013$\times E_{\rm GRB}/10^{51} \, {\rm erg}$ per century, respectively). Furthermore, even if such astrophysical sources are able to inject a large enough flux of positrons, they would be expected to produce a signal that traces both the disk and bulge components of our Galaxy. Since the observed emission is roughly spherically symmetric, astrophysical sources also require a network of coherent magnetic fields or some other mechanism to transport the positrons from the disk to throughout the volume of the bulge before they annihilate~\cite{Prantzos:2005pz}.

Motivated by these difficulties involved in explaining the observed 511 keV emission with astrophysical sources, it has been suggested that this signal might instead originate from annihilating~\cite{511dark} (also decaying~\cite{Hooper:2004qf} or exciting~\cite{Weiner}) dark matter particles. Due to the narrow width observed in the 511 keV line, the positrons must be injected with energies less than a few MeV~\cite{beacom}. For this reason, the corresponding dark matter particles must also have an MeV-scale mass.

Interestingly, it has been shown that if dark matter consists of $\sim$MeV mass particles which annihilate primarily to $e^+ e^-$ through a $P$-wave process (such that $\sigma v \propto v^2$), then its annihilations will inject the required rate of positrons into the Galactic Bulge, and simulaneously be produced in the early universe with approximately the measured dark matter abundance~\cite{511dark}. For example, if the dark matter consists of a stable, MeV mass scalar or fermion which annihilates to $e^+e^-$ through the exchange of a new light gauge boson, the relic abundance and 511 keV flux can be easily accomodated~\cite{511dark,scalar,fayet}.

Despite the success of this simple phenomenological picture, no realistic particle physics model containing such particles has appeared in the literature to date. Furthermore, there appears to be common opinion within the particle physics community is that it is very difficult to build a realistic and natural model that possesses the features required of this scenario. In an effort to challenge this conclusion we propose here a natural supersymmetric model which contains a stable, MeV-scale dark matter candidate with the features required to generate the observed 511 keV emission from the Galactic Bulge.

The cross section for the annihilation of Majorana dark matter particles, $\tilde{X}$, through the exchange of a vector boson, $U$, is given by~\cite{scalar,drees}:
\begin{eqnarray}
\sigma v &=& \frac{g^2_{UXX} \sqrt{1-4 m^2_f/s}}{12 \pi s [(s-m^2_U)^2 + \Gamma^2_U m^2_U]} \,  \bigg[ \{s (g^2_{Uf_Lf_L}+g^2_{Uf_Rf_R})\\ \nonumber 
&+&m^2_f (6g_{Uf_L f_L}g_{Uf_Rf_R}-g^2_{Uf_Lf_L}-g^2_{Uf_Rf_R})\} \, (s - 4 m^2_{\tilde{X}}) \\ \nonumber
&+& \, \bigg(\frac{m_f m_{\tilde{X}}}{m^2_U}\bigg)^2 (3g^2_{Uf_Lf_L}+3g^2_{Uf_Rf_R} - 6g_{Uf_L f_L}g_{Uf_Rf_R})\bigg].
\end{eqnarray}
Here $g_{UXX}$, $g_{Uf_Lf_L}$ and $g_{Uf_Rf_R}$ are the mediator's coupling to dark matter, left-handed fermions and right-handed fermions, respectively. The first part of this expression (the term in the $\{\,\}$ brackets) vanishes in the low velocity limit, being the result of a $P$-wave amplitude~\cite{fayet}. The second term, however, provides a contribution to the low velocity cross section. The expression for the case a scalar dark matter candidate is the same, but without the second term.

In order for MeV-scale Majorana dark matter particles to annihilate with the rate required to generate the observed 511 keV flux and the measured dark matter abundance, the low velocity cross section must be suppressed. Furthermore, the product of the couplings of the mediating particle to electrons and the dark matter must be: $g_{UXX} \times g_{Uee} \sim 10^{-5}$-$10^{-7} \times (m_U/10\,{\rm MeV})^2$~\cite{drees,scalar,fayet,Hooper:2007tu}. Measurements of the electron's magnetic moment and other constraints further require  $g_{UXX} \gsim g_{Uee}$. Such constraints are satisfied if, for example, the gauge coupling of the mediator to the dark matter is $g_{UXX} \sim {\cal O}(1)$, while the coupling of the mediator to electrons is $g_{Uee} \sim 10^{-5} \times (m_U/10\,{\rm MeV})^2$. 

We motivate our model building by the observation that the ratio between the MeV masses required in this model and the electroweak scale is also $\sim 10^{-5}$. Therefore, we can hope to construct a natural mechanism by which the fields of the hidden sector (including the dark matter candidate and mediator) have their MeV-scale masses generated through radiative corrections suppressed by their $\sim 10^{-5}$ couplings to the visible sector supersymmetry breaking masses.

With this goal in mind, we set out to construct a model including a stable dark matter candidate with an MeV-scale mass which annihilates to electrons through an MeV-scale mediator with an ${\cal O}(10^{-5})$ effective coupling. We begin with a minimal model consisting of one chiral superfield, $\Xi$, and one vector superfield, $U$. Together, these fields constitute a hidden sector.  The superfield, $\Xi$, plays a dual role in our model. In particular, the scalar component of $\Xi$, which we denote as $X$, breaks the $U(1)_h$ symmetry associated with $U$, while its fermionic component is the lightest supersymmetric particle, is thus stable by the virtue of R-parity conservation, and thus constitutes our dark matter candidate.  We discuss briefly near the end of the paper how he ${\cal O}(10^{-5})$ effective coupling of $U$ to the Standard Model fields can be generated; for now we take this small coupling to be the single input of the model which generates the mass scale for the MeV dark matter and mediator.

\begin{figure}
\resizebox{9.0cm}{!}{\includegraphics{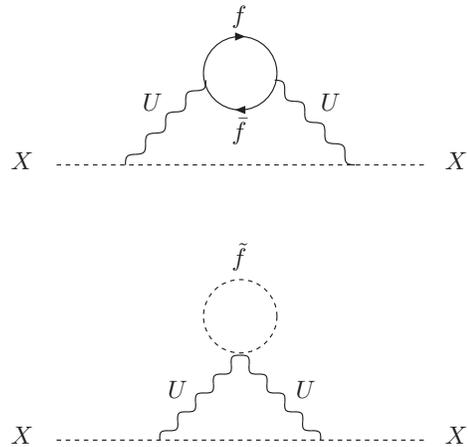}} \\
\caption{Examples of two loop diagrams which generate the mass of the scalar, $X$. The solid and dotted lines in the loop represent Standard Model fermions and their scalar superpartners.}
\label{fig2}
\end{figure}

In order to break the $U(1)_h$ symmetry and give the mediator $U$ a mass, the scalar component, $X$, of the chiral superfield, $\Xi$, which is charged under $U(1)_h$, must get a vacuum expectation value (vev), in a manner exactly analogous to the Higgs mechanism.  In order for the vev to be stable and non-zero, the scalar potential must have both a negative mass squared, which we will generate through radiative corrections, and a $|X|^4$ contribution, which comes from SUSY-preserving $D$-terms.  The $D$-term gives a contribution to the potential of $\frac{1}{2} h_X^2 g^2 \left| X \right|^4$, where $h_X$ is the charge of $X$ under $U(1)_h$.
The negative mass squared which breaks the symmetry results from two loop diagrams analogous to gauge mediated SUSY breaking~\cite{nelson}, except in this case the Standard Model fields and their superpartners run in the loop, instead of messenger fields (see Fig.~\ref{fig2}).   The radiative mass contribution to $X$ from a single scalar in the loop is given approximately by (see Ref.~\cite{poppitz}):
\be
m_{X, rad}^2 \approx -\frac{g_{UXX}^2 g_{Uff}^2}{64 \pi^4} m_S^2 \log\frac{\Lambda_{UV}^2}{m_S^2},
\label{radmass}
\ee 

where $m_S$ is the mass of the MSSM scalar in the loop and $\Lambda_
{UV}$ is the scale at which the SUSY breaking mass for the scalar is
generated which, for concreteness, we take to be $10^9$ GeV.  $g_
{UXX} = g h_X$ is the coupling of the mediator $U$ to $X$, and $g_
{Uff} = g h_f$ is the coupling of $U$ is Standard Model fermions, where $h_f$ is
the charge of $f$ under $U(1)_h$.  The fewest constraints on the
model occur when $g_{UXX} \sim {\cal O}(1)$ and $g_{Uff} \sim 10^{-5}
$, so that a very small charge, $h_f \sim 10^{-4}$-$10^{-5}$, is necessary to
make the model phenomenologically feasible (see Ref.~\cite{Hooper:2007tu}
for a summary of constraints).  We return to the question of how
such small charges may be generated later.

The radiative mass is negative since, unlike most models of gauge mediation, the messenger supertrace is not zero. 
To get the total contribution to the $X$ mass one must include all the MSSM scalars and fermions in the loop which have a non-zero $h_{Uff}$ coupling.  In general, we can write this as 
\be
m_{X, rad}^2 \approx -6 \MeV^2 \left(\frac{g_{UXX}^2 \, \sum_f g_{Uff}^2 m_{\tilde{f}}^2}{10^{-10} (4 \mbox{ TeV}^2)}\right).
\ee
The value of the $X$ mass thus depends on the spectrum of the MSSM states, which is model dependent, though the heaviest scalar (typically a $\tilde{t}$) will contribute most.  Once stabilized by the $D$-term, the vev of $X$ breaks the $U(1)_h$ symmetry, and gives the $U$ gauge boson a mass.  In particular, $\langle X \rangle^2 =| m_{X,rad}^2|/2 g^2 h_X^2 $ so that 
$m_U^2 = 2 g^2 h_X^2 \langle X \rangle^2 = |m_{X,rad}^2|$
and  $m_X^2 =  2 |m_{X,rad}^2|$.  The fermionic components of $U$ and $\Xi$ get masses through $\tilde{U}-\tilde{X}$ mixing of size $m_{\tilde{U}} = m_{\tilde{X}} = \sqrt{2} g h_X \langle X \rangle$, which is degenerate in mass with the mediator.

This basic scenario demonstrates that one can easily build a model in which a hidden sector scalar with a natural MeV-scale mass exists and generates masses for all of the fields in that sector (including the gauge field and all fermionic superpartners) which are also MeV-scale.  We have not, however, taken care to build a sector which is anomaly free.  The simplest possibility, adding a $\Xi_c$ chiral superfield with opposite charge to $\Xi$ under $U(1)_h$, will introduce a D-flat direction in the potential.  The D-flat direction may be stabilized with the addition of a singlet $S$ (as in the NMSSM) with superpotential
\be
W = \lambda S \Xi \Xi_c,
\ee
giving a scalar potential which is
\begin{eqnarray}
V  & = & \frac{1}{2} \left( m_{X,rad}^2 |X|^2 + m_{X_c,rad}^2|X_c|^2 \right)  \\ \nonumber
&+& \lambda^2\left( |S|^2 (|X|^2 + |X_c|^2) +| X^2| |X_c^2| \right) + \frac{g^2}{2} |X^2 - X_c^2|^2.
\end{eqnarray}
The minima of $X$, $X_c$ derived from the scalar potential are $\langle X_c,X \rangle^2 = -m_{X,rad}^2/\lambda^2$, and the mass eigenstates of the $X$,~$X_c$ are $m_{X_1}^2 = 2 \lambda^2 \langle X \rangle^2$ and $m_{X_2}^2 = (4 g^2 - 2 \lambda^2) \langle X \rangle^2$.  The gauge field gets a mass $m_U^2 = 4 g^2 \la X \ra^2$ from the vev of the $X$, $X_c$ fields.  The fermionic components become massive through mixing.  In the $(\tilde{X},\tilde{X}_c,\tilde{U},\tilde{S})$ basis, the mixing matrix is
 \be
{\cal M} =
\left( 
\begin{array}{cccc}
0 & 0 & a \langle X \rangle & \lambda \la X \ra \\
0  & 0 &  -a \langle X \rangle & \lambda \la X \ra \\
a \langle X \rangle & -a \langle X \rangle & 0 & 0 \\
\lambda \la X \ra & \lambda \la X \ra & 0 & 0 \\
\end{array} 
\right),
\ee  
where $a = \sqrt{2} g$ for $h_X = 1$, giving mass eigenstates $2 g \la X \ra$ (two) and $\sqrt{2} \lambda \la X \ra$ (two).  Since we require $2 g^2 > \lambda^2$ to obtain symmetry breaking, we can see that one of the scalar eigenstates and two of the fermion eigenstates are typically the lightest states of the theory.  When $X$,~$X_c$ obtain vevs all U(1)'s are broken so that the scalars are unstable, and the fermion is the dark matter candidate in this theory.

We now comment on how a small charge $h_f \sim 10^{-4}$ may be
generated.  The most straightforward way is through kinetic mixing.
The $U$ gauge boson, for example, may mix with the standard model
hypercharge through a term in the Lagrangian $\chi U^{\mu \nu} F_{\mu
\nu}$.  In this case, the Standard Model $Z$ gets slightly modified couplings (as
in Ref.~\cite{Ng}), where the deviation $\sim \chi$ is small enough to be
consistent with precision electroweak constraints.  Kinetic mixings
as small as $10^{-4}$ can be naturally generated \cite{Dienes}.  A
light, MeV-scale mass, $Z'$ (which is mostly $U$) couples the hidden sector
$X$ to Standard Model fields carrying hypercharge, giving an effective charge
$h_f \sim Y_f \chi g'/g$, where $g'$ is the Standard Model hypercharge gauge
coupling.  While this scenario may be attractive for generating small
charges using only Standard Model hypercharge and a small kinetic mixing piece $
\chi$, it does present an obstacle for shielding the hidden sector
sufficiently from MSSM SUSY breaking.  We have assumed in the model
presented above that the SUSY breaking is communicated through MSSM
fields in the two loop graphs, Fig.~\ref{fig2}, generating $m_{X,rad}^2$.  If
gauge mediatiion is the dominant source of SUSY breaking for the
MSSM, and the small charges result from $U$ mixing with Standard Model
hypercharge, one may worry that messenger particles running in the
loop would generate a larger positive contribution for $m_{X,rad}^2
$.  We thus must either assume that the dark matter sector does not
mix with $U(1)_Y$ and that the small charge $h_f$ is generated
through some other means, or that it is sequestered in some other way
from MSSM gauge mediated SUSY breaking.

Although we have described here only one model which gives all the relevant features, there are in principle many sets of chiral superfields one could add which would generate an anomaly free sector.  Others may be explored, and they may have a rich phenomenology.  The model we have given here demonstrates proof of principle for natural models with scalars with masses $\ll \mbox{ TeV}$.

As an additional note, the presence of this hidden sector makes the MSSM LSP unstable to decays to hidden sector particles.  This leads to the appearance of R-parity violation and, as a result, key missing energy signals used to search for supersymmetry at colliders may be reduced. This is similar to the behavior discussed within the context of ``Hidden Valley'' models~\cite{hidden}.  It will be interesting to investigate further the impact of such a model of MeV dark matter on collider phenomenology.

To summarize, motivated by the observation of 511 keV emission from the Galactic Bulge, we have presented a simple and natural supersymmetric model that contains a viable MeV dark matter candidate. In this setup, the MeV mass of the dark matter particle is generated naturally from radiative corrections through its small couplings ($\sim 10^{-5}$) to the Minimal Supersymmetric Standard Model.   The MeV mass of the gauge particle which mediates the interactions between the dark matter and Standard Model also naturally results.  Such models of hidden sector dark matter are novel and natural extensions of the Minimal Supersymmetric Standard Model which result in unique cosmology, such as the 511 keV signal, and non-standard supersymmetric phenomenology at the LHC.

\medskip

This work has been supported by the US Department of Energy, including grant DE-FG02-95ER40896, and by NASA grant NAG5-10842.  We thank Dan Chung, Bogdan Dobrescu, Lisa Everett, Roni Harnik, Tom McElmurry, Ann Nelson, Neal Weiner, and especially David E. Kaplan and Frank Petriello for discussions.

\end{document}